\documentclass[epj]{webofc}%
\usepackage{amsmath}
\usepackage[varg]{txfonts}
\usepackage{amsfonts}
\usepackage{amssymb}
\usepackage{graphicx}%
\woctitle{MESON2016 - the 14$^\textrm{th}$ International Workshop on Meson Production, Properties and Interaction}
\begin{document}
\selectlanguage{english}
\title{Modelling glueballs}

\author{Francesco Giacosa\inst{1,2}\fnsep\thanks{\email{fgiacosa@ujk.edu.pl}} 
}

%insert email only for speaker/presenter

\institute{
Institute of Physics, Jan Kochanowski University,
ul. Swietokrzyska 15, Kielce, Poland
\and
Institute for Theoretical Physics, J.W. Goethe University,
Max-von-Laue-Str. 1, Frankfurt am Main, Germany}

\abstract{Glueballs are predicted in various theoretical approaches of QCD (most notably lattice QCD),
but their experimental verification is still missing.
In the low-energy sector some promising candidate for the scalar glueball exist, and some
(less clear) candidates for the tensor and pseudoscalar glueballs were also proposed. Yet, for heavier gluonic states
there is much work to be done both from the experimental and theoretical points of view.
In these proceedings, we briefly review the current status of research of glueballs and discuss
future developments.}

\maketitle

\section{Introduction}

The fundamental particles of Quantum Chromodynamics (QCD) are quarks and
gluons. Both are colored: quarks in the fundamental representation of the
color group $SU_{c}(3)$ (red, green, blue), gluons in the adjoint
representation (color-anticolor, minus the white configuration). The
fundamental principle on which QCD is built is the invariance under local
color transformations.

Color is not directly seen: confinement implies that the physical states
emerging from QCD are `white'. For instance all conventional quark-antiquark
($\bar{q}q$) states have the white wave function$\sqrt{1/3}(\bar{R}R+\bar
{B}B+\bar{G}G).$ Such states constitute the majority of the mesonic resonances
listed in the PDG \cite{pdg}, see also the review papers \cite{amsler} and the
predictions of the quark model \cite{isgur}.

From the early days of QCD \cite{chodos,jj,ryzak} it was clear that bound
states made solely of gluons, called glueballs, might exist. In fact, gluons
interact strongly with themselves. The existence of glueballs was also
predicted by various methods, most notably by lattice QCD, e.g. Refs.
\cite{morningstar,mainlattice,gregory}, both in the quenched and unquenched
approximations. Yet, while their existence seems compelling from the
theoretical point of view, up to now no resonance was found which can be
\textit{unambiguously} recognized as a predominantly glueball state. While in
the low-energy sector (below 2.6 GeV) some candidate exist, in the high-energy
sector no candidate is known. Experimental searches at low energies in the
very soon upcoming experiments GlueX \cite{gluex} and CLAS12 \cite{clas12} at
Jefferson Lab and at high energy at the ongoing BESIII \cite{bes,bespsg} and at the future PANDA \cite{panda} experiments are expected to
improve our understanding.

The theoretical and experimental work on glueballs has been huge: up to now
(status on 26/9/2015), there are 1404 papers which contain the world glueball
in the title (for reviews, see Refs. \cite{ochs,vento,crede}). In these
proceedings we present some recent developments on this fascinating and still
unsolved piece of QCD.

\section{Existence and masses of glueballs}

Photons do not interact with each other at tree-level. (A quartic photon
interaction emerges through a fermionic loop, whose amplitude is suppressed by
$\alpha^{2}$.) Gluons are completely different: they shine, already at the
leading order, in their own light. This fact, together with confinement,
naturally leads to the idea that bound states made of solely gluons should exist.

The early studies of glueballs were performed within bag models, e.g. Refs.
\cite{chodos,jj,ryzak}. In particular, in\ Ref. \cite{ryzak} various
microscopic currents were introduced and a glueball spectrum was shown. The
lightest states were the scalar and tensor glueballs (at about 1 GeV),
followed by pseudoscalar and pseudotensor ones.

The development of lattice QCD allowed to perform quantitative and model
independent studies of the QCD spectrum. Already in 1999 a complete spectrum
of glueballs (in the quenched approximations, i.e. without quarks) was
presented \cite{morningstar}. The lightest state is a scalar glueball with a
mass of about 1.7 GeV, followed by the tensor and the pseudoscalar states.
This result has been confirmed by numerous and more recent lattice
calculations, see Ref. \cite{mainlattice}, which is currently cited in the PDG
in the review on the quark model \cite{pdg}.

\begin{center}
\textbf{Table 1}: Central values of glueball masses from lattice (from
\cite{mainlattice}).%

\begin{tabular}
[c]{|c|c|}\hline
$J^{PC}$ & Value [GeV]\\\hline
$0^{++}$ & $\,1.70$\\\hline
$2^{++}$ & $2.39$\\\hline
$0^{-+}$ & $2.55$\\\hline
$1^{-+}$ & $2.96$\\\hline
$2^{-+}$ & $3.04$\\\hline
$3^{+-}$ & $3.60$\\\hline
$3^{++}$ & $3.66$\\\hline
$1^{--}$ & $3.81$\\\hline
$2^{--}$ & $4.0$\\\hline
$3^{--}$ & $4.19$\\\hline
$2^{+-}$ & $4.22$\\\hline
$0^{+-}$ & $4.77$\\\hline
\end{tabular}

\end{center}

Calculations within unquenched lattice QCD (i.e., with quark fluctuations)
basically confirmed the same trend \cite{gregory}, in turn, meaning that the
mixing and the decays of glueballs should not be too large. This is indeed a
very good information for model builders and experimental searchers.

While lattice QCD is the best theoretical proof of the existence of glueballs
and the most reliable calculation of masses, other approaches were also used
by theoreticians: QCD sum rules \cite{qcdsumrules}, Hamiltonian QCD
\cite{hamqcd}, flux-tube model \cite{fluxtube}, anti De-Sitter approaches
\cite{ads}, and Bethe-Salpeter equations \cite{bs}. All of them find glueballs
and the scalar state is the lightest.

In conclusion, there is nowadays a great confidence about the existence of
glueballs and about the qualitative form of the spectrum. Nevertheless, the
identification of glueballs has still to come.

\section{Decays of glueballs}

\textbf{Large-}$N_{c}$

According to the famous large-$N_{c}$ limit \cite{thooft} (simplifications
occurs when the number of colors $N_{c}$ is artificially increased to large
values), glueballs' masses scale with $N_{c}^{0},$ just as $\bar{q}q$ masses.
The decay of glueballs into mesons scales as $N_{c}^{-2},$ which is even more
suppressed than regular $\bar{q}q$ states (that scale as $N_{c}^{-1})$. It is
then expected that glueballs are narrow. This theoretical consideration is
particularly important for future PANDA project \cite{panda}, which will
search for glueballs between 2.2-5 GeV. Namely, only if glueballs are
sufficiently narrow, they can be discovered experimentally.

\textbf{Flavour and chiral blindness}

Glueballs are flavour-invariant, hence they should decay in a flavour-blind
way. For instance, for a glueball decaying into two pseudoscalar states and
neglecting phase space, one obtains the ratios $\pi\pi:KK:\eta\eta
:\eta^{\prime}\eta^{\prime}:\eta\eta^{\prime}=3:4:1:1:0$. In addition, the
decays of a glueball are also chirally invariant, since it couples with the
same strength to all chiral partners (such as $\rho\rho$ and $a_{1}%
(1230)a_{1}(1230)$)$.$

\textbf{Scalar glueball}

The ground-state scalar glueball is undoubtedly the most studied gluonium. In
the literature, many different scenarios concerning the identification of the
scalar glueball in the realm of scalar states listed in the PDG were proposed.
In most cases, the result was that the biggest gluonic amount is either in
$f_{0}(1500)$ or in $f_{0}(1710)$, e.g. Refs.
\cite{close,kirk,weingarten,longglueball,cheng,constdecay,gutsche,stani,chenlattice,rebhan}%
. A very short summary of the historical development is the following: in the
pioneering work of Amsler and Close \cite{close}, later on confirmed by Close
and Kirk \cite{kirk}, the largest gluonic amount sits in $f_{0}(1500).$ This
conclusion was reached analyzing the decays of the three resonances
$f_{0}(1370),$ $f_{0}(1500),$ and $f_{0}(1710)$ into two pseudoscalar states
using a $^{3}P_{0}$ approach. On the other hand, Lee and Weingarten
\cite{weingarten} used a lattice QCD approach to study the mass of the scalar
glueball and its couplings to pions and kaons: the outcome was that
$f_{0}(1710)$ is mostly gluonic. In Ref. \cite{longglueball}, Giacosa et al.
used an hadronic model inspired by ChPT in which also the other members of the
scalar nonet were included, $K_{0}^{\ast}(1430)$ and $a_{0}(1450).$ The fit to
all decays showed the existence of two solutions, one in which $f_{0}(1500)$
is predominantly a glueball, and on in which $f_{0}(1710)$ is such. Shortly
after, Cheng et al \cite{cheng} also found a phenomenological solution in
which $f_{0}(1710)$ is predominantly a glueball. Various other studies were
performed which involved constituents quarks and gluons, e.g. Ref.
\cite{constdecay}, or which involved the decay of the $j/\psi$ meson, e.g. Ref.
\cite{gutsche}.

The scalar glueball is also linked to the anomalous breaking of dilatation
symmetry (at the composite level, a dilaton/glueball field is introduced
\cite{migdal,salo}). In\ Ref. \cite{ellis} a peculiar fact was shown.\ Using
the dilaton potential from\ Refs. \cite{migdal,salo}, the decay of the
glueball into pions turned out to be about $4$ GeV, hence definitely too
large to be detected. The numerical value is obtained by assuming that the
dilaton saturates the gluon condensate \cite{gluoncondensat}. If this were
true, large-$N_{c}$ would badly fail in the scalar sector and one could never
find such a broad glueball.\ 

As discussed in Ref. \cite{stani}, the determination of the parameters of the
dilaton potential through the gluon's condensate is not necessarily true. More
in detail, in Ref. \cite{stani} the glueball was studied within the so-called
extended Linear Sigma Model (eLSM). This is an hadronic model based on chiral
symmetry and dilatation invariance together with their explicit and
spontaneous breaking. The eLSM, first developed for two flavours \cite{nf2},
has shown to be capable to describe masses and decays mesons up to $1.7$ GeV,
as the three-flavour study of Ref. \cite{dick} shows. The glueball as a
dilaton is naturally included in this model.\ Quite remarkably, there is only
a solution within the eLSM: $f_{0}(1710)$ is mostly gluonic. This result is in
agreement with the original claim of Ref. \cite{weingarten}, but also with the
recent lattice study of Ref. \cite{chenlattice}, in which the decay
$j/\psi\rightarrow\gamma G$ is numerically evaluated. Moreover, the very same
conclusion has been reached in Ref. \cite{rebhan} by using an approach based
on the AdS/QCD correspondence. Future information form the GlueX experiment is
expected \cite{gutschenew}.

In conclusion, while a final assignment cannot yet be done, there is mounting
evidence from different directions that $f_{0}(1710)$ is mostly gluonic.$\ $

\textbf{Tensor glueball}

According to lattice, the tensor glueball has a mass of about 2.2 GeV (it is
second lightest). In Ref. \cite{anisovich} it was pointed out that the
resonance $f_{J}(2220)$ does not lie on the Regge trajectories. Moreover, the
state is very narrow, the $\pi\pi/KK$ ratio is in agreement with flavour
blindness \cite{tensor}, and no $\gamma\gamma$ decay was seen. A necessary
improvement would be the experimental assessment of this candidate. In
particular, it is not yet clear if $J$ is $2$ or $4.$ Nevertheless, this is a
promising starting point for future studies (for instance, employing the eLSM).

\textbf{Pseudoscalar glueball}

The pseudoscalar glueball has been also investigated in a variety of
scenarios, see Ref. \cite{masoni} for a review. One has investigated the
gluonic content of the resonance $\eta^{\prime},$ e.g.\ Refs. \cite{kloenew}
and \cite{escribano}. In various other works, e.g. Ref. \cite{tichy}, the
pseudoscalar glueball was assigned to the resonance $\eta(1405),$ while
$\eta(1295)$ and $\eta(1475)$ are $\bar{q}q$ states. Such a scenario is
controversial for two reasons: (i) At present, it is not clear if $\eta(1405)$
and $\eta(1475)$ are two independent states. (ii) The mass of the pseudoscalar
glueball as predicted by lattice QCD is about 2.6 GeV, i.e. 1 GeV heavier.

In Ref. \cite{psg} the eLSM has been used to study the decays of an
hypothetical pseudoscalar glueball (linked to the chiral anomaly
\cite{salopsg}) with a mass of about 2.6 GeV, in agreement with lattice. The
outcome was that the decay channels into $KK\pi$ and $\eta\pi\pi$ are
dominant, while $\pi\pi\pi$ should vanish. A possible experimental candidate
is the state $X(2370)$ measured by BES \cite{bespsg}, yet future measurements
on its decay rates are needed.

\textbf{ Other glueballs}

The other glueballs listed in\ Table 1 need further studies. Very recently,
two steps have been performed: (i) in\ Ref. \cite{ptg} the decays of a
pseudotensor glueball has been studied in a flavour-invariant hadronic model:
sizable decay into $K_{2}^{\ast}(1430)K$ and $a_{2}(1320)\pi$ and vanishing
decays into $\rho\pi$ are predicted. (ii) The decays of a vector glueball in a
fully chirally invariant approach (using the eLSM) have been investigated
in\ Ref. \cite{vg}: a sizable decay into $\omega\pi\pi$ (both direct and
indirect through $b_{1}\pi$) and into $\pi KK^{\ast}(892)$ are expected to be
the main signal of a vector gluonium. Such simple predictions may help future
experimental searches.

\section{Conclusions}

Glueballs are expected to exist but were not yet found in experiments. While
GlueX and CLAS12 can help our understanding in the light sector, BESIII and,
in the future, PANDA can search for glueballs in the heavy sector. Definitely,
more work is needed: predictions about the decay channels of glueballs might
be particularly helpful in the process of identifications of possible
candidates. The aim is to close the gap between a basic theoretical
expectation of QCD and the present experimental status.

\textbf{Acknowledgments}: Support from the Polish National Science Centre NCN
through the OPUS project nr. 2015/17/B/ST2/01625 is acknowledged.

\end{document}